\documentclass[rnote]{aa} 
\usepackage{graphicx}
\usepackage{txfonts}
\usepackage{natbib}
\begin{document}
   \title{On the Relation of Hard X-ray Peak Flux and Outburst Waiting Time in the Black Hole Transient GX~339-4}

   \author{Y. X. Wu\inst{1}\fnmsep\inst{2} \and W. Yu\inst{1} \and Z. Yan\inst{1}
   \and L. Sun\inst{1} \and T. P. Li\inst{2}
          }

   \institute{Key Laboratory for Research in Galaxies and Cosmology and Shanghai Astronomical
Observatory , 80 Nandan Road, Shanghai, China\\
              \email{wenfei@shao.ac.cn}
         \and
             Department of Engineering Physics \& Center for Astrophysics, Tsinghua University, Beijing, China\\
       \email{wuyx@mails.thu.edu.cn}
             }

   \date{}


  \abstract
   {}
   {In this work we re-investigated the empirical relation between
the hard X-ray peak flux and the outburst waiting time found
previously in the black hole transient GX~339-4. We tested the
relation using the observed hard X-ray peak flux of the 2007
outburst of GX~339-4, clarified issues about faint flares, and
estimated the lower limit of hard X-ray peak flux for the next
outburst. }
   {We included Swift/BAT data obtained in the past four years.
Together with the CGRO/BATSE and RXTE/HEXTE light curves, the
observations used in this work cover a period of 18 years. }
   {The observation of the 2007 outburst confirms the empirical
relation discovered before. This strengthens the apparent link
between the mass in the accretion disk and the peak luminosity of
the brightest hard state that the black hole transient can reach. We
also show that faint flares with peak fluxes smaller than
about 0.12~crab do not affect the empirical relation. We predict that the hard X-ray peak flux of the next outburst should be larger than 0.65~crab,
which will make it at least the second brightest in the hard X-ray since 1991.}
   {}

   \titlerunning{On the Relation of Hard X-ray Peak Flux and Waiting Time in GX~339-4}
   \authorrunning{Wu et al.}
   \maketitle

%

\section{INTRODUCTION}
GX~339-4 is a black hole transient discovered more than 30 years
ago. It has a mass function of $5.8~M_{\odot}$, a low mass companion
star and a distance of $\gtrsim 7$~kpc \citep{Mar73, Hyn03,
Sha01,Zdz04}. It is one of the black hole transients with the most
frequent outbursts \citep{Kon02,Zdz04}.  \citet{Yu07} analyzed the
long-term observations of GX~339-4 made by the Burst and Transient
Source Experiment (BATSE) on board the {\it Compton Gamma-Ray
Observatory} (CGRO) and the {\it Rossi X-ray Timing Explorer} (RXTE)
since May 31, 1991 until May 23, 2005. They found a nearly linear
relation between the peak flux of the low/hard (LH) spectral state
that occurs at the beginning of an outburst and the outburst waiting
time defined based on the hard X-ray flux peaks. The empirical
relation indicates a link between the brightest LH state that the
source can reach and the mass stored in the accretion disk before an
outburst starts.

After then the source underwent an outburst in 2007. The 2007
outburst and any future outbursts can be used to test and refine the
empirical relation. Here we show that the hard X-ray peak flux of
the 2007 outburst falls right on the empirical relation obtained by
\citet{Yu07}, proving that the empirical relation indeed holds. By
including the most recent monitoring observations with the Swift/BAT
in the past four years, we re-examine the empirical relation and
make a prediction for the hard X-ray peak flux of the next bright
outburst for a given waiting time. We also clarify issues related
to faint flares that have been seen in the recent past.

\section{OBSERVATION AND DATA ANALYSIS}
We made use of observations performed with BATSE (20--160~keV)
covering from May 31, 1991 to May 25, 2000, HEXTE (20--250~keV)
covering from January 6, 1996 to January 2, 2006, as in
\citet{Yu07}, and recent monitoring results of Swift/BAT that are
publicly available (15--50~keV) covering from February 13, 2005 to
August 31, 2009. The BATSE data were obtained in crab unit. The
fluxes of the Crab were 305~counts~s$^{-1}$ and
0.228~counts~s$^{-1}$~cm$^{-2}$ for HEXTE and BAT respectively.
These values were used to convert the source fluxes into the unit of
crab. Following the previous study \citep{Yu07}, the light curves
were rebinned to a time resolution of 10~days. It is worth noting
that the X-ray fluxes quoted below all correspond to 10-day
averages, including those obtained in the empirical relation and the
predicted fluxes.

\begin{figure}
   \resizebox{\hsize}{!}{\includegraphics{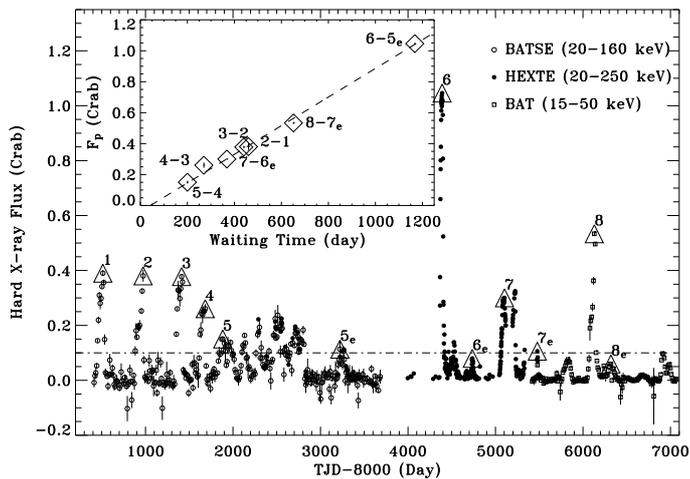}}
   \caption{The long-term hard X-ray light curves of
GX~339-4 from the observations with CGRO/BATSE (empty circles),
RXTE/HEXTE (filled circles) and Swift/BAT (squares). The inset panel
is the relation between the LH state peak flux and the waiting time
following Yu et al. (2007). The dashed-dotted line indicates the
flux level of 0.1~crab, under which X-ray flares appears not to
affect the empirical relation. The triangles indicate the LH state
peaks used to calculate the waiting times. The dashed line in the
inset panel shows the best-fitting linear model.}
   \label{fig_pkwt}
\end{figure}

\begin{figure}
   \resizebox{\hsize}{!}{\includegraphics{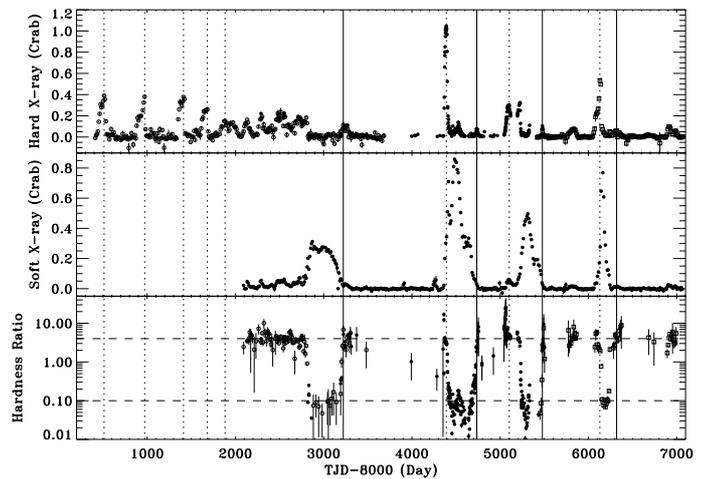}}
   \caption{Top: the long-term hard X-ray light curves of
GX~339-4 from the observations with CGRO/BATSE (empty circles),
RXTE/HEXTE (filled circles) and Swift/BAT (squares). Middle: the
long-term soft X-ray light curve of the RXTE/ASM (2--10~keV).
Bottom: The hardness ratios between the hard X-ray fluxes and the
ASM fluxes: BATSE/ASM (empty circles), HEXTE/ASM (filled circles)
and BAT/ASM (squares). The initial hard X-ray peaks and the hard
X-ray peaks at the end of the HS-to-LH state transition in the
outburst decays are marked with dotted lines and solid lines,
respectively. The HS and LH state hardness ratio thresholds were set
to 4 and 0.1 respectively, as indicated by the dashed lines.}
   \label{fig_hr}
\end{figure}

The combined BATSE, HEXTE and BAT light curves are shown in
Fig~\ref{fig_pkwt}. The triangles marked with 1--8 indicate the
initial hard X-ray peaks during the rising phases of the outburst
1--8, and those with $\rm 5_e$--$\rm 8_e$ indicate the ending hard
X-ray peaks during the decay phases of the outburst 5--8. Outburst
1-7 were studied in \citet{Yu07}. Outburst 8 is the 2007 outburst
that occurred after the empirical relation was obtained. The waiting
time of outburst 8 is determined in the same way as in the previous
study, i.e., the time separation between the peaks $\rm 7_e$ and 8
and the peak $\rm 7_e$ is the hard X-ray peak associated with the
HS-to-LH transition. In order to show how the peaks are chosen, we
also plotted the soft X-ray light curves obtained with the RXTE/ASM
and the hardness ratios between the ASM and the BATSE or HEXTE or
BAT fluxes in Fig~\ref{fig_hr}. This explicitly shows that the hard X-ray peaks at the end of outbursts correspond to the HS-to-LH state
transitions. The initial hard X-ray peak, on the other hand, is
normally the first prominent one during the initial LH state. Due to
the hysteresis effect of spectral state transitions \citep{Miy95},
the source would have very low luminosity after the HS-to-LH
transition during the outburst decay. We took the hard X-ray peak corresponding to the HS-to-LH state transition such as peak $\rm 7_e$, as the end of the previous
outburst, i.e., the starting time to calculate the waiting time of
the following outburst (see the definition of waiting time in
\citealt{Yu07}).

Due to the relatively low sensitivity of BATSE, flares with 10-day
averaged peak flux at or below about 0.1~crab could not be
identified as individual outbursts. It is therefore worth noting that the current empirical relation is determined
based on outbursts with hard X-ray peak fluxes above about 0.2~crab. In recent years with more
sensitive observations of Swift/BAT, we have observed several faint
flares in this source. These flares would not have been clearly seen
in the BATSE 10-day averaged light curve and would not have been
taken as single outbursts if BATSE had operated. Therefore we
ignored these flares although they were clearly seen with Swift/BAT.
We will discuss the faint flares later on.

We found that the data point of outburst 8 follows the empirical
relation reported in \citet{Yu07}, as shown in the inset panel of
Fig~\ref{fig_pkwt}. The deviation from the empirical relation is
only -0.034~crab. The linear Pearson's correlation coefficient for
all the 7 data points is 0.997, again indicating a nearly linear
relation between the hard X-ray peak flux $\rm F_p$ and the waiting time
$\rm T_w$.

A linear fit to this relation gives $\rm F_p=(9.25\pm0.06)\times
10^{-4}{\rm T_w}-(0.039\pm0.005)$, where $\rm F_p$ is in unit of
crab and $\rm T_w$ in units of days. This updated relation is almost
identical to the one reported in \citet{Yu07}. The intrinsic
scattering of the data is 0.014~crab, which defines a
$\pm$0.014~crab bound of the linear relation. The intercept of
the best-fitting linear model on the waiting time axis is $\rm
T_w=42$~days when $\rm F_p=0$~crab. Considering the intrinsic
scattering and the model uncertainty, we obtained an intercept
$\rm T_w= 42\pm 20$ days. This means that the hard X-ray peak of any
outburst should be at least $42\pm 20$ days after the end of the
previous outburst, which is determined as the hard X-ray peak
corresponding to the HS-to-LH transition.

\begin{figure}
   \resizebox{\hsize}{!}{\includegraphics{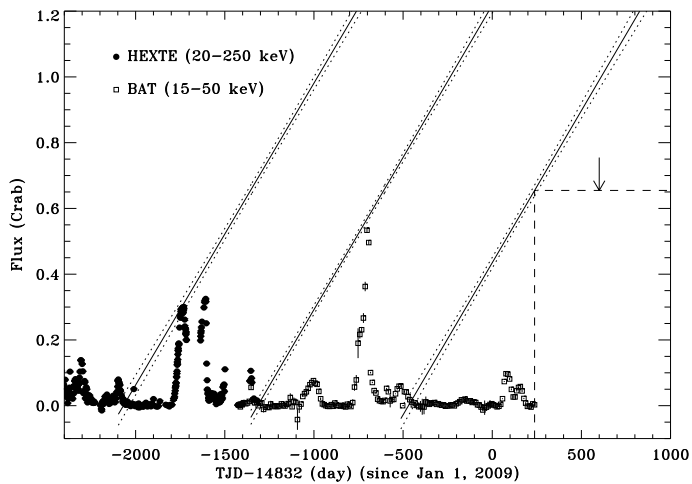}}
   \caption{Prediction of the hard X-ray peak flux for
next bright outburst based on the updated empirical relation. The dashed line indicates the regime excluded for the next bright outburst. The arrow indicates the lower limit of the hard X-ray flux of the next bright outburst (in comparison to those flares fainter than 0.12 crab) is about 0.65 crab estimated on August 31, 2009 (MJD 55074), the later afterwards the higher the lower limit will be. The ``predictions'' for
the 2004 and 2007 outburst (outburst 7 and 8 in
Figure~\ref{fig_pkwt}) based on the respective empirical relations
derived with data before 2004 and 2007 outburst are shown as well.
Data are from HEXTE (filled circles) and BAT (squares). The solid
lines are the predicted values of the peak fluxes. The dotted lines
show the corresponding prediction bounds at a level of 95\%. }
   \label{fig_pred}
\end{figure}

The refined empirical relation enables us to approximately estimate
the hard X-ray peak flux (10-day average) for the next bright
outburst in GX~339-4. The updated relation gives the peak flux of
the next bright outburst as $\rm F_{p,n}=9.25\times10^{-4}~({\rm
Day_{09}}+{\rm T_{rise}})+0.44$~crab, where $\rm Day_{09}$ is the
number of days in 2009 when a future outburst starts and ${\rm
T_{rise}}$ is the rise time in unit of day for the next outburst to
reach its initial hard X-ray peak. The hard X-ray peak flux can be
predicted almost as soon as the next outburst occurs because
the rise time is nearly a small constant compared with the waiting
time. The source has remained inactive for about 750~days since
the end of the 2007 outburst. This gives that the hard X-ray
peak flux of the next outburst should be at least 0.65~crab
(Fig~\ref{fig_pred}), making it the
second brightest outburst since 1991, brighter than all the outbursts
except outburst 6. Again notice that only for an
outburst brighter than about 0.12~crab can such a prediction be made
based on the empirical relation. We have shown that the empirical
relation holds if faint hard X-ray flares are ignored. For example
the flare of about 0.08~crab in March 2006 does not affect the peak
flux of the 2007 outburst. This suggests that the flare of about
0.1~crab in March 2009 will not affect the hard
X-ray peak flux of next bright outburst significantly.

The negligible effect of the faint flares on the empirical relation
is also consistent with the consideration of the actual value range
of $\rm T_w$. The intersection of the best-fitting linear empirical
relation on the time axis indicates that the hard X-ray peak of a
major, bright outburst must occur more than $42\pm20$~days after the
hard X-ray peak during the decay phase of the previous outburst.
However as discussed in \citet{Yu07}, the sum of the decay time of
the LH state in the previous outburst and the rise time of the LH
state in the next outburst is normally about 100--150~days.
Therefore in reality the minimal $\rm T_w$ for bright outbursts, as
defined by \citet{Yu07}, would be 100--150 days. This corresponds to
$\rm F_{p}$ in the range of $\sim0.04-0.12$~crab, which indicates a
lower limit of $\rm F_{p}$ for any outburst that should be
considered in the empirical relation. This might suggest that after an outburst,
GX~339-4 can subsequently rise up to $\sim0.12$~crab without returning quiescence. Because of their low luminosities, the faint flares correspond to only a small portion of
the mass in the disk. This consistently explains why using the
empirical relation without including the faint flares can we estimate
the hard X-ray peak flux of a bright outburst -- the indicator of
the disk mass.

In order to get an idea on how good the estimation or prediction is,
we also ``predict'' the hard X-ray peak flux for 2004 and 2007
outbursts with the data before the 2004 and the 2007 outburst respectively,
and then compared the ``predictions'' with the observations
(Fig~\ref{fig_pred}). We then studied the deviations of the
predicted values from the actual observed peak fluxes during the
2004 outburst and the 2007 outburst. The deviations are -0.012~crab
and -0.034~crab, or 3.8\% and 6.4\%, respectively. Considering that
the 10-day time binning would bring uncertainties, these predictions
are extraordinarily good. The prediction made for the next bright
outburst should have a similar accuracy. The hard X-ray peak of the
next outburst should fall on the prediction in Fig~\ref{fig_pred}
with a lower limit around 0.65 crab, which is the predicted hard X-ray peak flux of an outburst if it happened at present (around MJD 55074).

\section{DISCUSSION}
We included recent hard X-ray monitoring observations of GX 339-4
with Swift/BAT in addition to CGRO/BATSE and RXTE/HEXTE
observations. We have analyzed the X-ray observations of GX~339-4 in
the past 18 years following \citet{Yu07} and re-examined the
empirical relation between the hard X-ray peak flux and the outburst
waiting time during bright outbursts found by \citet{Yu07}. We found
that the hard X-ray peak flux of the 2007 outburst follows the
empirical relation determined with observations before 2007 very
well. We checked the potential influence of faint flares on the
empirical relation. The empirical relation was determined based on
the observations of bright outbursts, not including those faint
flares below about 0.12~crab. The actually minimal waiting time
required for an outburst to occur consistently explains that there exists a lower limit of peak flux
for the outburst studied here. A refined relation between the hard
X-ray peak flux and the waiting time in the past 18 years has been
obtained. Based on this relation, we can estimate the hard X-ray
peak flux for the next bright outburst as soon as it starts.
It has been 750 days since the end of the most recent bright outburst. Based on this, we
predict that the hard X-ray peak flux should be no less than 0.65~crab.

One may think that during different outbursts the properties of the
accretion flow are different, such that the radiation efficiencies
differ for different outbursts while the actual mass accreted are
about the same. This is not the case. The correlation between
the hard X-ray peak flux and the peak flux of the corresponding HS
state is found to hold for individual black hole binaries and
neutron star low mass X-ray binaries \citep{YKF04,YD07,YY09}. Given
that the neutron star has a hard surface, the observed X-ray flux
from the neutron star system should in general reflect the
instantaneous mass accretion rate. Therefore outbursts with
different flux amplitude in neutron star systems should correspond
to different mass accretion rate. Because the black hole systems
fall on the same correlation track as those,
the mass accretion rates should be different when GX~339-4 reaches
the hard X-ray peaks during outbursts of different amplitudes.

The empirical relation, confirmed by the BAT observations of the
2007 outburst, provides strong evidence that there is a link between
the mass in the accretion disk and the brightest LH state that
GX~339-4 can reach. The mechanism behind this link is not clear.
But if the mass in the accretion disk is directly related to the
production of the hard X-ray flux, then a major portion of the disk
should be involved in generating the hard X-ray flux. Independent of
such accretion geometry considerations, \citet{YY09} have recently
performed a comprehensive study of spectral state transitions in
bright Galactic X-ray binaries. The results have confirmed the
correlation between LH-to-HS transition luminosity and the peak
luminosity of the following soft state shown in previous studies
\citep{YKF04,Yu07,YD07}, and provided strong evidence for that: a)
non-stationary accretion plays a dominant role in generating a bright
LH state and b) the rate-of-increase of the mass accretion rate can
be the dominant parameter determining spectral state transitions. The
empirical relation between the LH-to-HS state transition luminosity
and the peak luminosity of the following HS state and the empirical
relation studied in this paper connect the mass in the accretion
disk (the cause and initial condition) and the peak luminosity of
the hard state (the result) to the rate-of-increase of the mass
accretion rate. This then could be the indicator of the initial mass which influences the
overall development of the hard state and the soft state and the
transitions between the two. The empirical relation allows us to estimate the mass in
the accretion disk before an outburst in the special source GX~339-4.

The phenomenon reminds us of a storage mechanism that works behind. This may be relevant to the phenomenon seen in solar flares, known as avalanche processes \citep[e.g., ][]{LH91,Whe00}, of which the magnetic field plays the major role.

\begin{acknowledgements}
We would like to thank the CGRO/BATSE group at NASA Marshall Space
Flight Center and the RXTE and the Swift Guest Observer Facilities
at NASA Goddard Space Flight Center for providing monitoring
results. WY would like to thank Robert Fender for stimulating
discussions and hospitality and encouragement which speeded up this
work. WY also thank Tomaso Belloni for a careful check of the BAT
flux for the 2007 outburst of GX 339-4. This work was supported in
part by the National Natural Science Foundation of China (10773023,
10833002), the One Hundred Talents project of the Chinese Academy of
Sciences, the Shanghai Pujiang Program (08PJ14111), the National
Basic Research Program of China (973 project 2009CB824800), and the
starting funds at the Shanghai Astronomical Observatory. The study
has made use of data obtained through the High Energy Astrophysics
Science Archive Research Center Online Service, provided by the
NASA/Goddard Space Flight Center.
\end{acknowledgements}

\end{document}